\documentclass[aps,pre,tightenlines,onecolumn,noshowpacs,superscriptaddress,longbibliography]{revtex4-1} 

\usepackage{graphicx}% Include figure files
\usepackage{dcolumn}% Align table columns on decimal point
\usepackage{bm}% bold math
\usepackage[normalem]{ulem}
\usepackage[usenames,dvipsnames]{xcolor}
\usepackage{amssymb,amsmath,amsfonts}   % for math
\usepackage{siunitx}%SI units

\newcommand{\vcrm}[1]{\bm{#1}}
\newcommand{\hvcrm}[1]{\bm{\hat{#1}}}
\newcommand{\vc}[1]{\bm{#1}}
\newcommand{\hvc}[1]{\bm{\hat{#1}}}
\newcommand{\dd}{\mathrm{d}}

\newcommand{\kk}{\mathrm{k}_B}
\newcount\colveccount
\newcommand*\colvec[1]{
        \global\colveccount#1
        \begin{pmatrix} 
        \colvecnext
}
\def\colvecnext#1{
        #1
        \global\advance\colveccount-1
        \ifnum\colveccount>0
                \\
                \expandafter\colvecnext
        \else
                \end{pmatrix}
        \fi
}

\begin{document}

\title{Biopolymer dynamics driven by helical flagella}

\author{Andrew K Balin}
\affiliation{Rudolf Peierls Centre for Theoretical Physics, 1 Keble Road, University of Oxford, OX1 3NP, United Kingdom}
\author{Andreas Z\"ottl}
\affiliation{Rudolf Peierls Centre for Theoretical Physics, 1 Keble Road, University of Oxford, OX1 3NP, United Kingdom}
\author{Julia M.\ Yeomans}
\affiliation{Rudolf Peierls Centre for Theoretical Physics, 1 Keble Road, University of Oxford, OX1 3NP, United Kingdom}
\author{Tyler Shendruk}
\email{tshendruk@rockefeller.edu}
\affiliation{Rudolf Peierls Centre for Theoretical Physics, 1 Keble Road, University of Oxford, OX1 3NP, United Kingdom}
\affiliation{The Rockefeller University, 1230 York Avenue, New York, New York, 10021}

\date{\today}% It is always \today, today,
             %  but any date may be explicitly specified

\begin{abstract}
Microbial flagellates typically inhabit complex suspensions of polymeric material which can impact the swimming speed of motile microbes, filter-feeding of sessile cells, and the generation of biofilms. There is currently a need to better understand how the fundamental dynamics of polymers near active cells or flagella impacts these various phenomena, in particular the hydrodynamic and steric  influence of a rotating helical filament on suspended polymers. Our Stokesian dynamics simulations show that as a stationary rotating helix pumps fluid along its long axis, polymers migrate radially inwards while being elongated. We observe that the actuation of the helix tends to \emph{increase} the probability of finding polymeric material within its pervaded volume. This accumulation of polymers within the vicinity of the helix is stronger for longer polymers. We further analyse the stochastic work performed by the helix on the polymers and show that this quantity is positive on average and increases with polymer contour length.
\end{abstract}

\maketitle

\section{Introduction}
\begin{figure}[t!]
\centering
 \includegraphics[width=0.75\columnwidth]{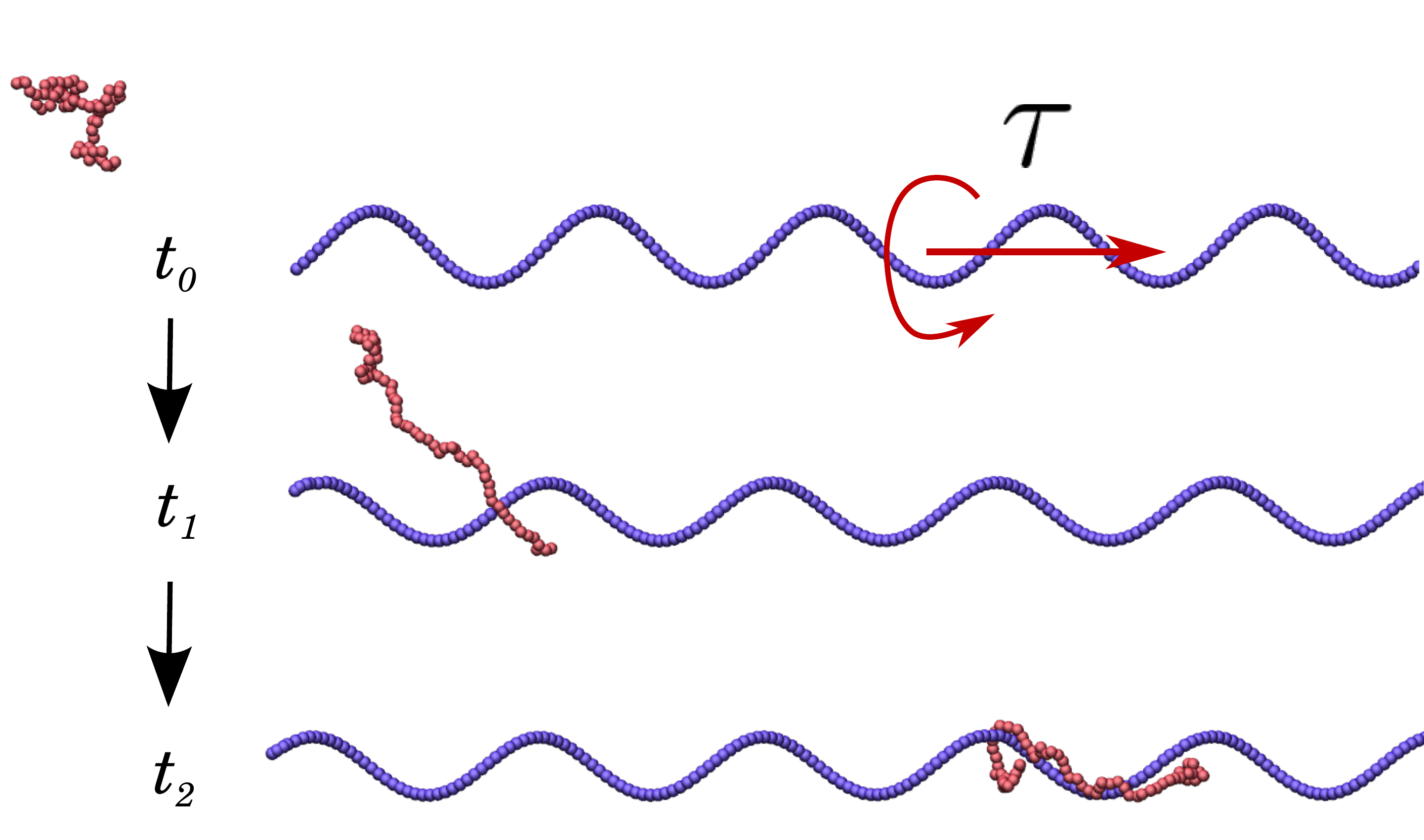}
  \caption{Snapshots from a single simulation of a rotating helix and an advected polymer taken at times $t_0 < t_1 < t_2$. A torque $\bm{\tau}$ is distributed uniformly on the helix and, as it rotates, it generates a pumping flow. The polymer is drawn in radially and axially towards the helix, stretching in the process due to the induced shear flow. Once captured it winds around the helix and is pumped axially in the direction of the average flow. \label{capture_single}}\label{snapshot}
\end{figure}
While the physics of swimming microbes in Newtonian viscous fluids has been well characterised, attention has recently turned towards understanding how active microorganisms behave in more biologically relevant media where the presence of large biopolymers, elastic filaments, or exopolymer secretions can dictate dynamics. 
In such complex fluids, motility enhancement \cite{Liu2011,Gagnon2013,Spagnolie2013,Thomases2014,Riley2014,Patteson2015}, as well as retardation \cite{Lauga2007,Fu2007,Spagnolie2013,Thomases2014,Shen2011,Zhu2012,Qin2015}, have both been reported for various biological swimmers. Theoretical studies exploring swimming in continuous viscoelastic media yield model- or parameter-dependent results \cite{Spagnolie2013,Thomases2014,Riley2014}. Invariably, these studies concentrate on continuum models of viscoelasticity, and as such cannot provide a full insight into the specific microscopic mechanisms of interaction between single polymers and the flagella of the swimmers.

Swimming dynamics may indeed be affected by such interactions due in part to the comparatively similar length and time scales of cells and biopolymeric material \emph{in vivo}. One hypothesis that has been applied to the swimming of \emph{E.\ coli} in dilute polymer solutions is that a timescale separation between the fast rotation of the bacterial flagellum and slow relaxation of the polymers effectively depletes the flagellum's local environment of polymeric material as it clears its surrounding volume. Hence, the flagellum experiences an effective viscosity that can be markedly different from that perceived by the more slowly counter-rotating cell body \cite{Martinez2014}. However, the microscopic assumptions underlying this hypothesis require more concrete justification. This raises the question of how large biopolymers interact with actively actuated filaments such as flagella on the single-polymer level. 

Additionally, there are separate phenomena in which the interactions between large polymers and active microorganisms are important but have yet to be studied on this microscopic level. For instance, various microbes have long been known to enhance filter-feeding by employing their flagella or cilia to generate feeding currents that carry detritus and nutrients toward the cell body \cite{higdon1979}. While motile neutrally buoyant planktonic bacteria \cite{marshall2013} swim force free \cite{Lauga2009,Elgeti2015}, these sessile microbes tether themselves in order to exert non-negligible net forces on the surrounding medium \cite{Pepper2009,Petroff2015}. Aggregations of such cells can collectively produce millimeter-scale fluid flows to actively combat variations in the nutrient concentrations \cite{Petroff2014}.
While hydrodynamic attraction of small nutrients has been considered \cite{Guasto2012} the interaction of macromolecular polymers with sessile flagellates has yet to be explored in detail. A question that remains open is whether the filter-feeding of large polymers progresses similarly to small tracer particles, or --as with swimming-- whether the dynamics of individual polymers enhances or reduces a tethered cell's ability to draw in polymeric material towards it.

In this work, we explore how macromolecular biopolymers hydrodynamically interact with an active helical pump that transports fluid by external actuation. A snapshot from our simulations is shown in Fig.\ \ref{capture_single}. We first construct a simple \emph{Stokesian dynamics} model of a rotating helix in a bulk fluid and justify the further use of this simplified model by comparing the flow field in the vicinity of the helix to that of a more specific biological model system comprised of a wall-tethered bacterium in a coarse-grained fluid. We show that the Stokesian dynamics scheme captures the relevant near-field physics of flow around a helix and proceed to study the dynamics of single polymers in its vicinity. Our results show that polymers are hydrodynamically drawn \emph{radially} inwards, and are elongated by the high shear rate. The polymers are then transported along the direction of the fluid flow while remaining elongated, and wrap around the helix before being deposited downstream. 

We also study the stochastic energetics of the helix to show that the helix actively performs work on the polymer, driving it to the higher free-energy state of elongation which we observe. This work is positive on average suggesting that the energy transferred to the polymer through hydrodynamic interactions is dissipated and not elastically transferred back to the helix.

\section{Methods}

\subsection{Stokesian dynamics simulation}

We employ a simulation scheme incorporating hydrodynamic interactions \cite{Ermak1978}, referred to as Stokesian dynamics (SD), to study the behaviour of a polymer in response to a steadily rotating helical filament. Both the polymer chain and the rotating helix are composed of sets of spherical particles whose positions $\{\bm{r}_i\}$ are updated according to Langevin dynamics due to forces $\{\bm{f}_i\}$ and thermal fluctuations $\{\bm{\xi}_i\}$:
\begin{eqnarray}
\vcrm{\dot{r}}_i & = &  \sum_j \Big( \kk T \frac{\partial \bm{\mu}_{ij}}{\partial \bm{r}_j}+ \bm{\mu}_{ij} \vcrm{f}_j\Big) + \bm{\xi}_i(t), \label{BD} \\
\langle \bm{\xi}_i (t)\bm{\xi}_j(t')\rangle & = & 2 \kk T \vc{\mu}_{ij} \delta(t-t'). \label{fluctuation}
\end{eqnarray}
Equations (\ref{BD}) are coupled in two ways: \emph{(i)} The total force $\bm{f}_i(\{\bm{r}_j\})$ acting on particle $i$ will in general contain a contribution due to pairwise interactions with nearby particles; \emph{(ii)} Hydrodynamic coupling between particles $i$ and $j$ is captured by the mobility tensor $\bm{\mu}_{ij}(\bm{r}_i,\bm{r}_j)$, which accounts for the advection of particle $i$ due to the flow-field created by forces $\bm{f}_j$ acting on particle $j$. The fluid medium is responsible for dissipating the momentum of the particles, demanding the fluctuating forces obey the fluctuation-dissipation relation [Eq.\ (\ref{fluctuation})], which correlates the fluctuations experienced by two widely separated particles.

We use the Rotne-Prager-Yamakawa (RPY) tensor for the mobility \cite{Rotne1969,Yamakawa1970},
\begin{align} \label{RPY}
\frac{\vc{\mu}_{ij}}{\mu_0} &= \begin{cases}
                                \frac{3a}{4r_{ij}}(\vcrm{I} + \vcrm{\hat{r}}_{ij}\vcrm{\hat{r}}_{ij}) +  \frac{a^3}{2r^3_{ij}}(\vcrm{I} - 3\vcrm{\hat{r}}_{ij}\vcrm{\hat{r}}_{ij}) & \text{for }r_{ij}\ge 2a \\
                                \Big(1 - \frac{9 r_{ij}}{32 a}\Big)\vcrm{I} + \frac{3}{32}\frac{r_{ij}}{a}\vcrm{\hat{r}}_{ij}\vcrm{\hat{r}}_{ij} &\text{otherwise},
                               \end{cases}
\end{align}
for $i\ne j$ and where $\vcrm{r}_{ij} = \vcrm{r}_j - \vcrm{r}_i$, $\mu_0=1/6\pi\eta a$ is the Stokes mobility of a sphere with radius $a$ immersed in a fluid with viscosity $\eta$, and $\vcrm{I}$ is the identity matrix. The self-mobility of particle $i$ is simply $\vc{\mu}_{ii} = \mu_0 \vcrm{I}$. The RPY tensor has the property that $\partial \bm{\mu}_{ij}/\partial \bm{r}_j = 0$, which simplifies the force balance equation (\ref{BD}).

\subsection{Polymer model}
All pairs of particles experience a mutual repulsion that acts over a characteristic length scale $\sigma$ and is given by the Weeks-Chandler-Andersen (WCA) potential,
\begin{equation} \label{repel}
\mathcal{H}_{repel} (\vcrm{r}_{ij}) = \begin{cases}
4\epsilon\Big[ \big(\frac{\sigma}{r_{ij}}\big)^{12} - \big(\frac{\sigma}{r_{ij}}\big)^6 \Big], & \text{if } r_{ij} < 2^{1/6}\sigma, \\
0, & \text{otherwise}.
\end{cases}
\end{equation}
The polymer is modelled as a chain of spherical beads, exerting pairwise attractive forces representing bonds between adjacent monomers. These bonding forces are calculated using the finitely-extensible non-linear elastic (FENE) potential \cite{Slater2009},
\begin{equation} \label{bond}
\mathcal{H}_{bond}(\vcrm{r}_{ij}) = -\frac{1}{2} k_{bond} r_0^2 \ln\Big[1-\Big(\frac{r_{ij}}{r_0}\Big)^2\Big],
\end{equation}
with Kremer-Grest parameters \cite{grest1986} $k_{bond}=30\epsilon/\sigma^2$ and $r_0=1.5\sigma$. We choose $\epsilon = \kk T$ for the characteristic strength of the potentials, $\sigma$ for the spatial unit, and $\sigma^2/\mu_0 \kk T$ for the temporal unit. This allows us to set $\kk T = 1$, $\sigma = 1$, $\mu_0=1$ hereafter. In these units, the bead diameter is $2^{1/6}$, hence $a=2^{-5/6}$. The conservative Hamiltonians $\mathcal{H}_{repel}$ and $\mathcal{H}_{bond}$ give rise to pairwise forces, $\bm{f}_i = -\bm{f}_{j} = -\nabla [\mathcal{H}_{repel}(\bm{r}_{ij})+\mathcal{H}_{bond}(\bm{r}_{ij})]$, which are the equal and opposite forces acting on particles $i$ and $j$.

For a 3-dimensional system of $N$ particles, Eqs.\ (\ref{BD}) can be rewritten in non-dimensional form as a $3N\times 3N$ matrix-vector difference equation involving time-step $\delta t$,
\begin{equation}
\delta\bm{r}=\ \bm{\mu} \bm{f} \delta t + \bm{b}  \delta\bm{w} \label{dr}
\end{equation}
where $\delta\bm{w}$ is a random Gaussian vector with the properties $\langle\delta\bm{w}\rangle=\bm{0}$ and $\langle\delta\bm{w}\delta\bm{w}\rangle=2\delta t\bm{I}$, and $\bm{b}$ is any matrix which satisfies $ \vcrm{b}\vcrm{b}^T=\vc{\mu} $. We find $\vcrm{b}$ by computing the Cholesky decomposition of $\vc{\mu}$ and note that Cholesky decomposition requires $\vc{\mu}$ to be positive-definite, which is ensured by the RPY tensor. Because the mobilities $\bm{\mu}_{ij}$ vary slowly with respect to the fastest timescales of the bond potentials, we update $\bm{\mu}$ once every 100 time-steps. 

\subsection{Rotating helix model}

The rotating helix comprises a set of particles whose individual positions are externally controlled by time-dependent forces that prescribe the shape, rotational frequency $\omega$ and translational velocity $v$ of the helix. This is done by applying a rectifying force that opposes displacements of a constituent particle from its prescribed location via a stronger harmonic potential,
\begin{eqnarray} \label{helix}
\mathcal{H}_{h}(\bm{r}_i,t) & = &  \frac{1}{2}k_{h} [ \bm{r}_i - \bm{r}_i^0 (t) ]^2, \\
\bm{r}_i^0 & = & \begin{pmatrix}
           R_0 \cos(\kappa z^0_i - \omega t) \\
           R_0 \sin(\kappa z^0_i - \omega t) \\
           z^0_i + vt
         \end{pmatrix}.
\end{eqnarray}
The prescribed positions $\{\bm{r}_i^0\}$ trace a helix along $\hvc{z}$, with an imposed radius $R_0$, and pitch \footnote{The pitch as usually defined as the length of a single complete turn, which in our notation is $2\pi/\kappa$} $\kappa$. For the helix potential strength we use $k_h=70\epsilon$. By applying the constraint that the target positions must be separated by diameter $2a$ in space, the required spacing in $z$ is approximated by $z^0_i = 2a/\sqrt{1+\kappa^2 R^2}$. We impose a constant angular rotation rate $\omega$ about the $\hvc{z}$-axis, and enforce stationarity by setting $v=0$ along the $\hvc{z}$-axis. This model is adequate for reproducing the hydrodynamics of a rigid helix. Alternatively, modelling the helix as a semi-flexible polymer would require solving bonding angle and dihedral angle potentials with large stiffnesses \cite{hu2015}, in turn requiring $\delta t$ to be many orders of magnitude below the shortest timescale of interest which for our purposes is the relaxation time $\tau_0 \sim a^2/\mu_0 \kk T$ of the SD beads. 

The helix particles (labelled by subscript $i$) are initialised in their target positions at $t=0$, and then the helix as a whole relaxes into a steady state after a short transient period. The steady state differs slightly from the target shape due to \emph{(i)} a viscous, drag-induced phase lag behind their target position, causing a shrinkage in their radial coordinate which can shown to be $R=R_0/\sqrt{1+(\mu_0\omega/k_{h})^2}$ when no hydrodynamic interactions are present and \emph{(ii)} an additional collective displacement along $\vcrm{z}$ due to chiral asymmetry in the hydrodynamic interactions $\sum_{ j ( \ne i)}\bm{\mu}_{ij}\cdot\bm{f}_j$ with the other helix particles. For $v=0$, these displacements $\delta z_i$ are counteracted by a net force on the helix $-\frac{\partial\mathcal{H}_{h}}{\partial  z} = -N_{h}k_{h}\delta z$ which is imparted to the fluid in the $\hvcrm{z}$-direction.

With this control over $\omega$ and $v$, we can drag, rotate or apply some combination of translation and rotation to the helix. The imposed rotation and translation implies an external axial force and torque via the linear mobility relation,
\begin{equation}\label{mobility}
\colvec{2}{v}{\omega} = \begin{pmatrix}
\mu_{tt} & \mu_{tr} \\
\mu_{rt} & \mu_{rr}
\end{pmatrix} \colvec{2}{f_z}{\tau_z}
\end{equation}
where the components $\mu_{tt}$, $\mu_{tr}=\mu_{rt}$, $\mu_{rr}$ are mobility coefficients for the helix as a whole, and not to be confused with the Rotne-Prager tensor $\bm{\mu}$. 

\subsection{Multiparticle collision dynamics simulation}

To verify that our SD helix model well approximates the near-field flow of a sessile flagellated microbe in the presence of environmental boundaries in a manner that does not depend significantly on cell body-induced or wall-induced hydrodynamic interactions, we construct a more biologically accurate but computationally costly model of a wall-tethered bacterium and simulate it using multiparticle collision dynamics (MPCD).

MPCD is a particle-based method to solve the Navier Stokes equations on a coarse-grained level where particle dynamics and interactions are solved in alternating streaming and collision steps \cite{Kapral2008,Gompper2009}. This method has been used successfully to model the hydrodynamics of microswimmers near surfaces \cite{Elgeti2009,Elgeti2010,Zottl2012a,Zottl2014,Schaar2015,Hu2015a,Theers2016,Eisenstecken2016}  
and wall-tethered flagella \cite{Reigh2012,Reigh2013}. The fluid is modelled by point-like effective particles with mass $m$ at positions $\bm{r}_i$ with velocities $\bm{v}_i$. 

In the streaming step, the fluid particles move ballistically for a time step $\delta t$,
and their positions are updated according to
\begin{equation}
\bm{r}_i(t+\delta t) = \bm{r}_i(t) + \bm{v}_i(t) \delta t.
\end{equation}
In the collision step, particles are sorted into cubic cells of side-length $h$, and all particles in a cell stochastically exchange momentum according to
\begin{equation}
\bm{v}_i(t+\delta t) = \bm{u}_{\xi} + \bm{v}_r + \bm{v}_P + \bm{v}_L,
\end{equation}where $\bm{u}_{\xi}$ is the mean velocity in the cell, $\bm{v}_r$ is a random velocity drawn from a Maxwell-Boltzmann distribution at temperature {$T$, and $\bm{v}_P$ and $\bm{v}_L$ are correction factors to  conserve momentum and angular momentum in the cell \cite{Noguchi2007}. All physical quantities are measured in units of cell length $h$, fluid mass $m$, and thermal energy $k_BT$. We use a time step $\delta t = 0.02 \sqrt{mh^2/k_BT}$ and a mean number of fluid particles per cell $\gamma=10$ resulting in high Schmidt and Mach numbers to reproduce near-incompressible viscous Newtonian flows at low Reynolds number \cite{Padding2006}.

Figure \ref{andreas}(d) includes a representation of the bacterium model we use. The cell body itself is modeled as a rigid superellipsoid \cite{Barr1981} defined by the surface$ \left[ \left( x/h_x \right)^{2/ \epsilon_2} + \left( y/h_y \right)^{2/ \epsilon_2}  \right]^{\epsilon_2 / \epsilon_1} + \left( z/h_z \right)^{2/ \epsilon_1} = 1$, where we use $h_x=h_y=2h$, $h_z=4h$, $\epsilon_1 = 0.5$ and  $\epsilon_2 = 1$. It is oriented perpendicular to a wall (located at $z=-50h$) and fixed at $(x_0,y_0,z_0) = (0,0,-46h)$. We add a second wall far away from the bacterium at $z=50h$, and use periodic boundary conditions in the $x$ and $y$ directions with $x,y, \in (-50h,50h)$.
The flagellum is modelled as a rigid helical polymer consisting of 57 point-like beads of mass $10m$ which are separated by $1h$. The helix is given by the curve
\begin{equation}
\bm{r}(z) = \left( R\left[1-e^{-\left(\kappa z/l_s\right)^2}\right]\cos(\kappa z), R\left[1-e^{-\left(\kappa z/l_s\right)^2}\right]\sin(\kappa z), z \right), 
\end{equation}
where we use $R=2h$ as the helix radius, pitch $\kappa= 1/h$, and $\kappa z$ is the phase of the helix measured from where it meets the body. A nonzero Higdon parameter \cite{Higdon1979b} $l_s=3$  ensures that the helix is attached at the center of the cell body surface.

To model no-slip boundary conditions at the walls and the cell surface a bounce-back rule for the fluid particles is used \cite{Gompper2009}. The transfer of momentum between the flagellum and the fluid is achieved by including the flagellum beads into the collision step \cite{Malevanets2000}. While the cell-body is kept fixed during the simulation, the helix is rotated with a constant angular velocity $\omega=(5.6\times10^{-4}) / \delta t$, and the surrounding flow field is measured for a time $t=10^5\delta t$ and averaged over 70 independent runs.

While MPCD provides the machinery for incorporating boundaries and cell-specific body geometry, measuring physical quantities requires averaging over a large number of identical simulations and therefore studying the details of polymer dynamics in a driven flow is a computationally demanding task. On the other hand, Stokesian dynamics is well suited for this kind of study so here we use a combination of these techniques to ensure agreement between a simple SD model of a flagellum and an MPCD model of a tethered cell, before studying the detailed polymer dynamics in the SD model only.

\section{Helix hydrodynamics}
\begin{figure}[t!]
\centering  \includegraphics[width=\columnwidth]{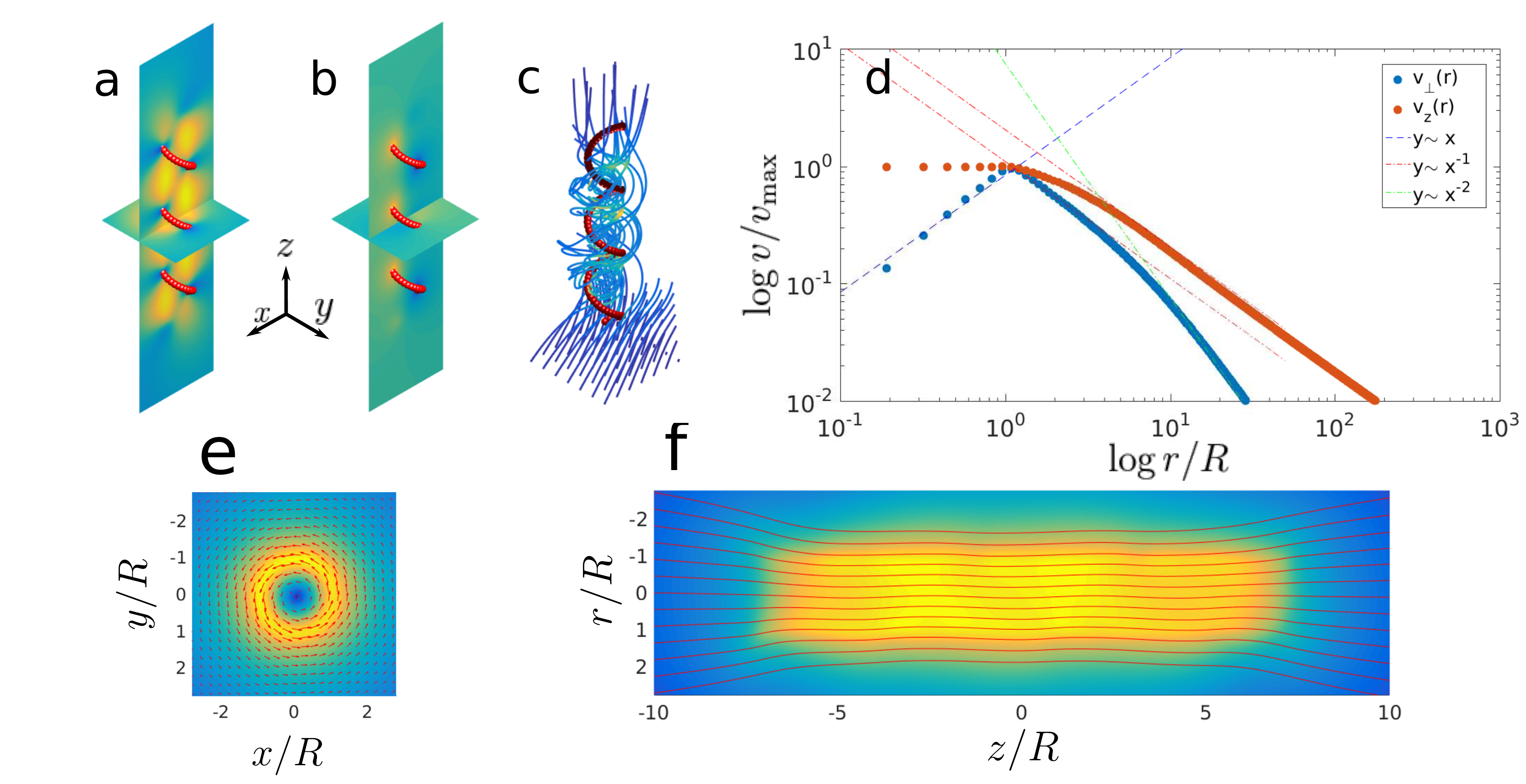}
  \caption{(a-b) Cross sections of the instantaneous flow field generated by a rotating helix, decomposed into components $v_z(\bm{r},t)$ and $v_\perp(\bm{r},t)$ respectively; (c) Streamlines of the full flow field $\bm{v}(\bm{r},t)$ shows both the net flow along $\hvc{z}$ and the chiral winding caused by the rotation of the helix; (d) Time-averaged fields as a function of radial distance from helix center-line, $\bar{v}_z(r,t)$ and $\bar{v}_\perp(r,t)$. Within the volume of the helical filament ($r<R$), the fluid rotates like a rigid body, $v_\perp \propto r$, and flows axially at a constant rate $v_r = \text{const}$; (e) Slice in the $xy$-plane of the mean rotational flow; (f) Slice in the $xz$-plane of the mean axial flow.\label{flow}}
\end{figure}
\subsection{Flow field generated by rotating helix in SD simulations}
We focus solely on a stationary helix ($v=0$) rotating at some angular speed $\omega$. This implies a non-zero force $f_z = -\frac{\mu_{tr}}{\mu_{tt}}\tau_z$ must be imparted to the fluid, and the rotating helix acts as a pump. This model evokes the microscopic experiments on tethered bacteria \cite{Yuan2010,Xing2006} as well as many scaled-up experiments of tethered flagella-like filaments \cite{kim2003,kim2004,yu2006,Coq2009,Zhong2013}. 

In the SD simulations of a helix on its own, we can evaluate the instantaneous flow field at any point $\bm{r}$ in space by summing the contributions that each particle in the simulation makes: $\bm{v}(\bm{r}) =\sum_{i}\bm{\mu}( \bm{r}-\bm{r}_{i})\cdot\bm{f}_i$ where $\bm{\mu}(\bm{r}')$ is given by Eq.\ (\ref{RPY}) with $\bm{r}_{ij}=\bm{r}'$. Figures \ref{flow}(a-c) offer a visualisation of the instantaneous flow field surrounding a rotating helix. In Fig.\ \ref{flow}(a), the axial component $v_z$ shows that the fluid is most strongly pumped in the interstitial volume of the helix, similar to the instantaneous axial flow field measured in experiments on a tethered rotating helix \cite{Zhong2013}. As the axial velocities of the helix beads are 0, $v_z$ must vanish at the helix surface. However, the transverse flow field $v_\perp$ is strongest at the helix surface as it must match the transverse velocity of the beads [Fig.\ \ref{flow}(b)]. An instantaneous snapshot of the streamlines originating from a square grid in the $xy$ plane beneath the helix gives a visual sense of the chiral nature of the flow field [Fig.\ \ref{flow}(c)]. 

By taking a time average $\bar{\bm{v}}(x,y,z)$ over a complete rotation of the helix, we can understand how the flow field varies in space in more detail. Figure \ref{flow}(d) shows how $\bar{v}_z$ and $\bar{v}_\perp$ decay as a function of radial distance $r$ from the $z$-axis along which the helix lies. In the far-field, we observe $\bar{v}_z \sim 1/r$, which is the characteristic scaling expected from a point-force (stokeslet) response of an unbounded fluid. This is as expected, since we must apply a force $f_z$ on the helix such that $\mu_{tt}f_z + \mu_{tr}\tau_z = 0$ by Eq.\ (\ref{mobility}) to ensure the rotating helix remains stationary ($v=0$). Hence, far away from the helix, the fluid responds as if subject to a point-force. 

The far-field scaling of the transverse velocity is the characteristic scaling for a rotlet, $\bar{v}_\perp \sim 1/r^2$, which we expect to dominate the far-field flow created by an external torque rotating a body immersed in the fluid. However, in the intermediate region ($R < r  < 10R$), the transverse forces on the beads on the near-side of the helix dominate over the oppositely directed forces on the far side, and therefore a stokeslet-like scaling $\bar{v}_\perp \sim 1/r$ is seen. 

Within the interior of the helix, we see interesting scaling properties $\bar{v}_z \sim \text{const}$, and $\bar{v}_\perp \sim r$, which shows that on average the fluid inside the helix rotates about $\hat{\bm{z}}$ and translates along $\hat{\bm{z}}$ as a rigid body ---though the instantaneous dynamics are more complicated.
Figures \ref{flow}(e-f) show the time averaged flow fields. The transverse flow field is strongest in the annular region occupied by the helical filament itself, while the axial flow field is uniformly strong across the whole volume.

\subsection{Friction tensor of helical filament in SD simulations}
\begin{figure}[t!]
\centering
  \includegraphics[width=0.8\columnwidth]{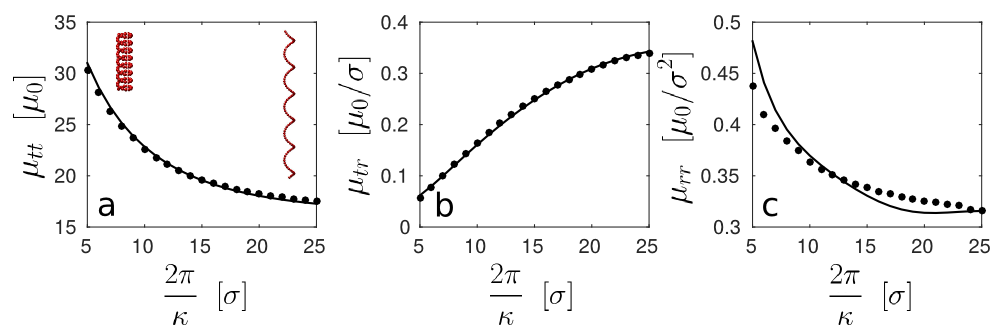}
  \caption{The three components of the helix friction tensor as a function of pitch length $2\pi/\kappa$. The data (black dots) are determined by measuring $f_z$, and $\tau_z$ separately as functions of $v$ and $\omega$ for different $\kappa$ in order to determine the matrix components $\mu_{tt}$, $\mu_{tr}$ and $\mu_{rr}$. The solid lines are calculated using analytical results from slender body theory (Eqs.\ (44) from Ref.\ \cite{Elgeti2015}). Increasing the pitch length decreases the isotropic components of the friction tensor, while increasing the coupling between rotation and translation. Inset: The helix shapes for the extremal choices of pitch.}
  \label{friction}
\end{figure}
In order to measure the helix mobility, we conducted SD simulations of a helix (with no polymer present) while linearly ramping up either the velocity or angular velocity, while keeping the other zero. In the first instance $(v=v_0 t/\mathcal{T},0)$ was imposed and in the second instance $(0,\omega=\omega_0 t/\mathcal{T})$ was imposed where $v_0$ and $\omega_0$ are the target final velocities and the length of the simulation, $\mathcal{T}$, was sufficiently long to ensure that the system remained in a quasi-steady state. In both cases, we measured $(f_z,\tau_z)$ in order to solve the linear system Eq.\ (\ref{mobility}). We conducted this for a range of $\kappa \in (\frac{2\pi}{25},\frac{2\pi}{5})$ to measure how the friction components changed as a function of helix shape. The functional dependence of these coefficients on $\kappa$ can be derived analytically using slender-body theory \cite{Lighthill1976,Rodenborn2013}. The SD simulation results are in good agreement with theoretical predictions \cite{Elgeti2015} as shown in Fig.\ \ref{friction}.

We observe that as the pitch length increases, $\mu_{tt}$ and $\mu_{rr}$ decrease, while $\mu_{tr}$ increases. This tells us that the coupling between axial force and rotation (or conversely between applied torque and resulting translational speed) increases as the pitch length is increased over the range shown. This behaviour can be understood intuitively by considering the limiting case of small pitch length $2\pi/\kappa \rightarrow \sigma$, in which the filament resembles a cylinder which by symmetry must have a totally decoupled mobility relation.  

The data for $\mu_{tt}$ and $\mu_{tr}$ fit very well to the slender body prediction. However, while $\mu_{rr}$ qualitatively displays a similar dependence on pitch length to the analytic prediction, it appears to systematically deviate from the theory. As previously discussed, the steady state shape of the helix realised in a simulation deviates slightly from its target shape [defined by Eq.\ (\ref{helix})]. This effect of this is generally small, but it appears that $\mu_{rr}$ is the more sensitive to this dynamic remodelling than the other components of the mobility matrix. However, in the remainder of this study, we will only conduct simulations in which the helix parameters remain constant and so this discrepancy in $\mu_{rr}$ as a function of $\kappa$ does not affect our findings. 

\subsection{Scope of rotating helix model}
Because our model helix remains in a fixed location in $\bm{\hat{z}}$, it is neither force-free nor torque-free, hence its flow field will differ to that generated by a swimming cell in the far-field. Artificial swimming magnetic ribbons \cite{Gao2014,Peyer2013} are arguably the most similar experimental realisation of our system due to the net torque they impart, but unless they are stalled by an external force, they remain force-free too. Hence, the force and torque nature of our model is more akin to systems in which flagellated cells are in some way stalled or tethered: e.g.\ by hydrodynamic accumulation at boundaries \cite{Spagnolie2012}, immobilisation on microscope slides \cite{Yuan2010,Xing2006}, or as part of their biological function \cite{Pepper2009,Petroff2015}. However in these cases, hydrodynamic interactions with the boundaries and cell bodies are a potential source of discrepancy between the results of our model and these experimental and biological systems.

\begin{figure}[t!]
\centering
  \includegraphics[width=0.8\columnwidth]{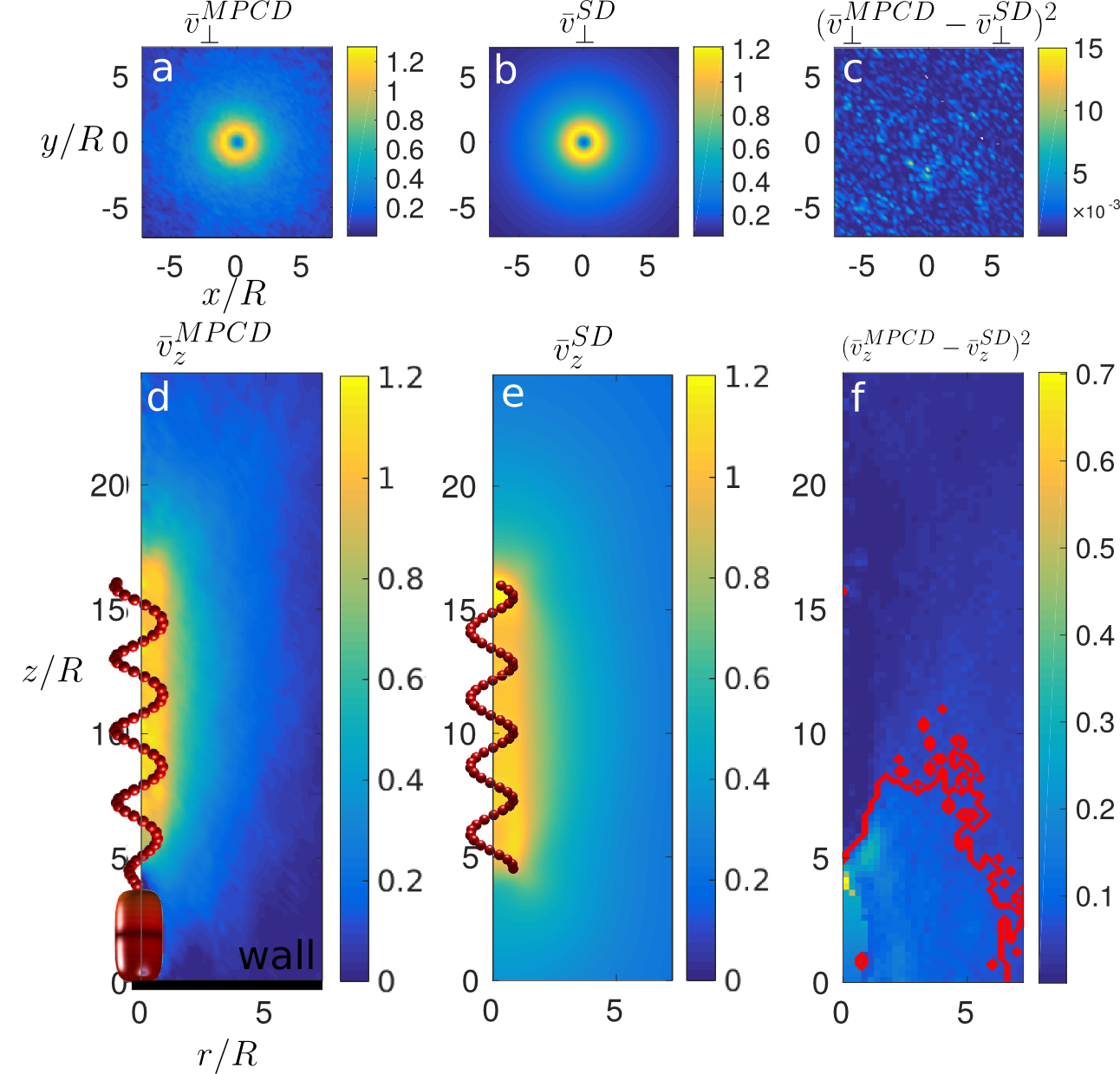}
  \caption{(a-c) Comparison of the time-averaged flow fields generated by (a) MCPD fluid simulations of a cell attached to a wall versus (b) SD simulations of a helix rotating in a bulk fluid computed in the $xy$-plane. These two measured fields differ only by Gaussian fluctuations inherent in the MPCD simulation as shown in (c). (d-f) Same as (a-c) except in $xz$-plane. (f) shows that there are systematic differences in the flow fields in this plane, mainly due to the presence of the wall. The contour-enclosed region in the lower quarter of the image represents where the flow field differs by $|\bar{v}_z^{MPCD}-\bar{v}_z^{SD}|>0.25$. Everywhere else, the fields behave similarly.
  }
  \label{andreas}
\end{figure}
To quantify this difference and determine whether our system is a sufficiently good model for studying the near-flagellum dynamics, we compared the flow fields measured from the SD simulations to MPCD simulations of a more experimentally realistic geometry with a cell body and neighbouring wall. In Fig.\ \ref{andreas}, we compared the time averaged flow fields $\bar{v}^{MPCD}$ and $\bar{v}^{SD}$ generated by the MPCD and SD simulations respectively. Both quantities were normalised by dividing by the mean velocity inside the volume occupied by the helix. By visual inspection, we observe that the coarse structure of the axial and transverse flows are qualitatively similar in both simulations.

To quantitatively compare the two fields, we analysed the square differences $(\bar{v}_{z,\perp}^{MPCD}-\bar{v}_{z,\perp}^{SD})^2$ to see how they decay relative to one another. In Fig.\ \ref{andreas}(c) we can see that $(\bar{v}_{\perp}^{MPCD}-\bar{v}_{\perp}^{SD})^2$ appears to have the structure of uniform noise. A Kolmogorov-Smirnov test on the data for the unsquared difference $\bar{v}_{\perp}^{MPCD}-\bar{v}_{\perp}^{SD}$ did not yield evidence for a non-Gaussian distribution, and so we conclude that $\bar{v}_{\perp}^{MPCD}$ differs from $\bar{v}_{\perp}^{SD}$ by the Gaussian noise present in the MPCD simulation only.

However, there are more significant differences in the radial flow fields due to the presence of the cell body and the wall in the MPCD simulations. The square difference $(\bar{v}_{z}^{MPCD}-\bar{v}_{z}^{SD})^2$, plotted in Fig.\ \ref{andreas}(f), reveals a systematic variation across the whole region that is roughly one order of magnitude greater than the noise in Fig.\ \ref{andreas}(c). The contour-enclosed area connected to the wall (at $z=0$) shows the region in which the flow fields differ by $|\bar{v}_z^{MPCD}-\bar{v}_z^{SD}|>0.25$. Evidently, the cell body and wall have some significant influence  on the flow in this region but not in the immediate vicinity of the model flagellum.

The fact that the radial fields are in better agreement than the axial fields can be understood by noting that $\bar{v}_\perp \sim 1/r^2$ decays over shorter distances than $\bar{v}_z \sim 1/r$, and hence the wall effects play a much larger role for the axial fields. 

\begin{figure*}[t!]
\centering
  \includegraphics[width=\columnwidth]{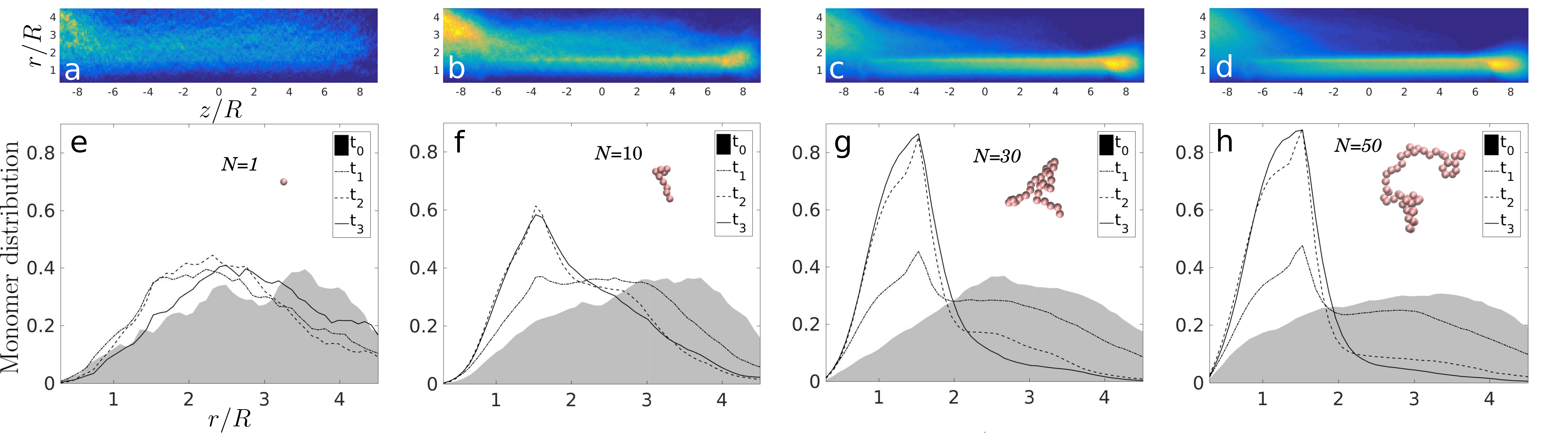}
  \caption{(a-d) Polymer distributions averaged over all simulations and over all time show the net behaviour of polymers being pumped in the positive-$z$ direction, given an initial distribution of polymers on a disc of radius $r=4R$, located at $z\approx-10R$ in cylindrical coordinates $(r,z)$. Each distribution is averaged over 200 simulations for polymers of size (a) $N=1$ (colloidal tracers), (b) $N=10$, (c) $N=30$, and (d) $N=50$. The net behaviour is a drift towards the right of each image, due to the helix (not shown) pumping the fluid. (e-h) The same data as (a-c) respectively, but plotted as distributions over $r$ only, with each curve representing contiguous quarter-intervals of the simulation time. In each case, the initial quarter is the shaded region, and we observe that for increasing polymer size, a strong tendency for the polymers to migrate inwards is observed. For larger polymers, this tendency is stronger and occurs faster.\label{capture}}
\end{figure*}
Since our main focus in this paper is the near-flagellum dynamics of polymers and helices interacting across length scales similar to and less than their own spatial dimensions, we take these MPCD results as evidence that far field effects (such as those generated by counter-rotating cell bodies, solid boundaries, and indeed other nearby swimmers or filaments) do not contribute appreciably to the dynamics of polymers sufficiently close to the helical filament. In this region, we expect the chiral, high-shear and geometry-specific flow of the helix to characterise the behaviour of a nearby polymer, and far-field effects due to boundaries or other bodies to be secondary. 

\section{Polymer capture}
The central result of this paper is that a rotating helix transports a polymer along with the fluid it pumps, but in such a way that a polymer initially on the outside of the helix is drawn inwards and `captured' by the helix. This is accompanied by an initial stretching out of the polymer as it migrates towards the helix, moving into a region of greater shear and greater flow as seen in Fig.\ \ref{capture_single}. As it is transported along the interior of the helix, it occasionally wraps around the helical filament and rotates along with it. An instance of such wrapping is observed in Fig.\ \ref{snapshot}. The polymer is deposited at the end of the helix, where a decaying axial current keeps moving it at a diminishing rate, while the lower shear results in the polymer collapsing back into its unstretched ground state. 

The polymer capture and transport is a stochastic, non-equilibrium transient process; however, by performing a large number of `scattering' simulations and averaging over these, we are able to quantify the typical nature of the interaction as a function of polymer size. We present batches of simulations for degrees of polymerisation (number of beads) $N_p=1, 10, 30, 50$ ---where a polymer of contour length 1 is simply a spherical monomer. In each of these simulations, we use helices with the same parameters: $N_h=200$ beads, $\kappa=2\pi/15\sigma$, $R=4\sigma$. The helix is centered along the $z$-axis, with its two ends located at $(z_0,z_{N_h-1}) = (-57\sigma, 57\sigma)$. In each simulation, one polymer is initialised by placing its first monomer randomly on a disc of radius $16\sigma$, located at $(z_0 - 30\sigma)$, then performing a self-avoiding random walk to build the polymer bead by bead. 

To illustrate how the ensembles of polymers evolve in time, we plot the average monomer distribution in cylindrical coordinates $(r,z)$ taken over all simulations and at all times. These are plotted for polymers of differing size in Fig.\ \ref{capture}(a-d). These images represent time- and ensemble-averaged 2D histograms of the snapshots shown in Fig.\ \ref{capture_single}. At the left edge of each of the images is the initial distribution of polymeric material which in each case is smeared rightward in time. The structure of this distribution gives a graphical indication that larger polymers [such as in Fig.\ \ref{capture}(c-d)] are much more strongly attracted to the helix than smaller polymers and are highly concentrated in the region $r<2R$. On the other hand, in Fig.\ \ref{capture}(a) this effect is barely observed for $N=1$ monomers, which are simply advected along the streamlines shown in Fig.\ \ref{flow}(f) like tracer particles that cannot cross streamlines --polymers, on the other hand are able to cross streamlines in shear flow \cite{graham11}, and in this case do so strongly in a non-trivial manner.

Particularly for high Weissenberg numbers, we observe hotspots in the distributions located at the downstream end of the helix. These indicate the accumulation of polymers in this location when they are deposited at the end of the helix and collapse back into their equilibrium conformation. Though they continue moving along the $z$-axis, they do so at a lower rate than when inside the helix. 

In Fig.\ \ref{capture}(e-h) we segment each simulation into four equal and contiguous intervals in time, and separately plot the marginal distributions over $r$ only (i.e.\ with the $z$-component integrated out) for each time. The shaded region represents the interval $t_0 = [0, \mathcal{T}/4]$, which is the first quarter of each simulation and closely approximates the initial distribution. These figures show how the initial distribution evolves with time for polymers of different size. For $N_p=30$ and $N_p=50$, the tendency to concentrate in and around the helix is markedly stronger than for shorter polymers.  This ensemble behaviour shows that the actuation of the helix is responsible for a large density fluctuation in the surrounding polymeric material that concentrates --rather than depletes-- the polymers in the immediately surrounding region. 

This implies that the free energy of the polymers must be actively driven away from what we would expect in equilibrium. We obtain an intuitive sense that this is occurring by considering the snapshots in Fig.\ \ref{capture_single}. Initially, the polymer is far away and its configuration is that of a self-avoiding random walk. However at intermediate times, the polymer is stretched out of equilibrium by the shear flow and is transported radially inwards as well as along $z$ until it strongly interacts with the helix, wrapping around it and continuing to move along $z$. In the vicinity of the helix, polymers lose their equilibrium conformation, and we observe features in their dynamics similar to those previously reported for polymers in shear flow due to a rotating microwire \cite{shendruk2016}. To gain further insight into the energetic interplay within our system, we analyse the stochastic fluctuations in work performed by the helix on the polymer.

\section{Fluctuating work}

Polymers tend to become stretched when immersed in a shear flow, and this agrees with the current SD simulations. However, because the flow field is generated by external forces acting on the helix, driving this stretching requires these external forces to be greater in order to do the extra work needed to drive the system out of equilibrium.

The work applied to the helix by the external forces, $w[t, \bm{r}(t)]$, is a fluctuating quantity which is a unique function for each realisation of a stochastic dynamical process. Work is performed either by the application of a non-conservative force, or by a time-varying potential, $\mathcal{H}(\bm{r},\lambda(t))$ with an external control parameter $\lambda(t)$. For the latter case, the work applied by a time $t$ is defined by \cite{Seifert2012}:  $w[t, \bm{r}(t)] = \int_0^t \dd t' \dot{\lambda} \partial \mathcal{H} / \partial \lambda$. The non-stochastic forces in our simulation are due to the potentials $\mathcal{H}_{repel}$, $\mathcal{H}_{bond}$ and $\mathcal{H}_{h}$ in Eqs.\ (\ref{repel},\ref{bond},\ref{helix}). Of these, only the forces acting on the helix due to $\mathcal{H}_h$ depend explicitly on time and it is these that are entirely responsible for the work done on the system. 

For each simulation, we calculate the incremental work performed by the helix at each time-step by,
\begin{equation} \label{work-eq}
\delta w  = \sum_i \bm{f}^{h}_i \cdot \delta \bm{r}_i^0,
\end{equation}
where $\bm{f}^{h}_i = -\nabla \mathcal{H}_h(\bm{r}_i, t)$ is time-varying force applied to particle $i$. Note the increment $\delta \bm{r}_i^{0}$ is the displacement of the bead target position, not the displacement of the bead itself. From these increments $\delta w$ we build up an accumulated work trajectory $w(t) = \sum_t \delta w$.

We expect there to be two contributions to the work: $w(t) = w_0(t) + w_{ex}(t)$. The dominant contribution $w_0(t)$ is the deterministic work done by the rigid helix on the viscous fluid, which is viscously dissipated. The second contribution is $w_{ex}(t)$, which is the stochastic excess work done on the polymer. By conducting simulations without a polymer, we can measure the dominant viscous contribution, $w_0(t)$, and use this to calculate the excess contribution in simulations that do contain a polymer: $w_{ex}(t) = w(t) - w_0(t)$.  

In Fig.\ \ref{work} we plot three ensembles of trajectories $w_{ex}(t)$ for the work done by a helix on three sizes of polymer: $N=(10, 20, 40)$. In each simulation, the polymer is initialised by a self-avoiding random walk starting at $\bm{r}_0=(0,15\sigma,(z_0 - 15)\sigma)$, where $z_0$ is the $z$-position of the negative-most particle of the helix which pumps fluid in the positive-$z$ direction. We measure $w_{ex}(t)$ for the entirety of each simulation over a time interval of 80 full rotations. For all simulations this is enough time to allow for the polymer to relax back to its equilibrium conformation after it has exited the positive-$z$ end of the helix.

These sets of trajectories offer another way to look at the stretching effect: a set of stochastic work trajectories $\{w_i[t,\bm{r}(t)]\}$ drive the polymer to a higher free energy state though in each instance requiring a different amount of work.} The colour of each curve denotes the radial point of closest approach of the center-of-mass position of the polymer --i.e.\ the minimum of $r_{cm}(t)/R$ over all $t$-- from which we can see that more work was done on polymers which migrated further in. This indicates that more work must be done on maintaining the stretched-out conformation which polymers adopt in the high-shear region of the helix core. 

\begin{figure}[t!]
\centering
  \includegraphics[width=0.8\columnwidth]{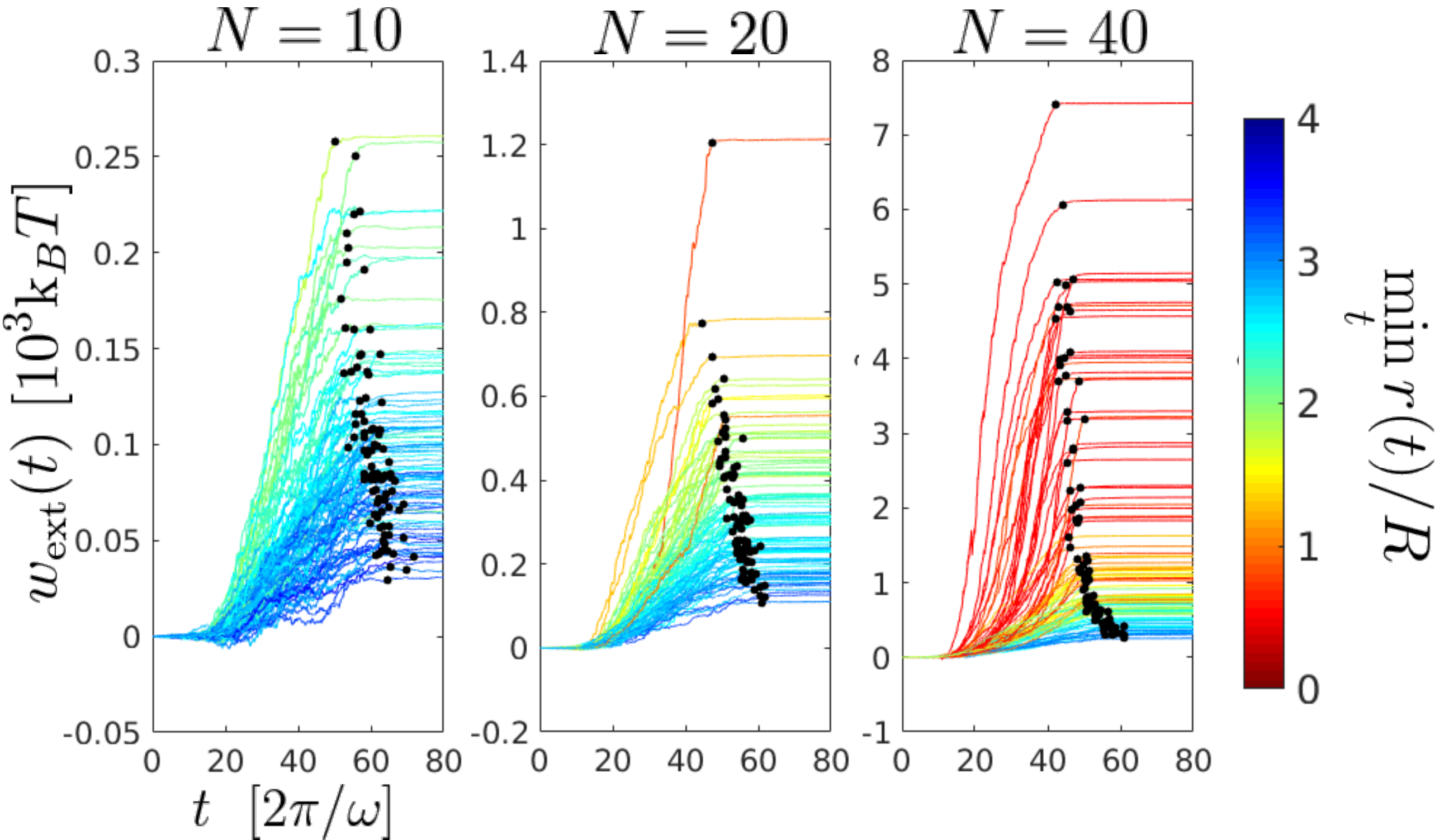}
  \caption{Stochastic work performed by a rotating helix in transporting polymers of size (a) $N=40$, (b) $N=20$, (c) $N=10$ in 100 experiments performed for each case. The trajectories are colour-coded by their point of closest approach, $\text{min}_t\ (r)$, to the central long axis of the helix. More work on average was performed to transport polymers that migrated nearer (red) the central axis than on polymers which failed to become captured (blue). This is because in the higher shear rate regions near the helix, polymers are stretched out into a higher free-energy state, which requires a net amount of work to be performed in order to maintain this state. The black dots correspond to the point at which the centre-of-mass of the polymer exits the negative end of the helix. Once this occurs, the polymers collapse and advect with the fluid; no more work is performed on them.}
  \label{work}
\end{figure}

The trajectories are non-monotonic, and some trajectories temporarily deviate into the negative work region, which is a hallmark of the thermodynamics of stochastic systems. We further notice that once the polymer has exited the helix, $w_{ex}(t)$ flattens dramatically. This corresponds to the observation that when the polymer is deposited at the rear end of the helix, it quickly collapses to its equilibrium configuration and highlights that the rotation of the helix principally impacts polymers in its immediate vicinity. Furthermore, the fact that $w_{ex}(t)$ does not overshoot its final value tells us that the helix does not regain any of the work it has supplied to the polymer when the polymer relaxes. This is in contrast to the observed enhancement of swimming due to the energetics of noiseless elastic surroundings \cite{Ledesma-Aguilar2013,Wrobel2016}. In such systems, elastic networks or tubes which a swimmer swims through store elastic energy and transfer this energy back to during relaxation. However, in our system, the heat bath to which the polymer is attached robs the helix of any such energy `storage' mechanism.

In general, the work excess term is typically smaller than the viscous term by $\sim4$ orders of magnitude. However, this is for a solitary polymer in the vicinity of the helix. In a suspension of polymers the excess work takes the form of a sum over the work performed on each polymer and so will be proportional to the local density of the solution at least in the dilute limit where polymer-polymer interactions can be ignored. As we have shown separately, the effect of the helix is to increase the local density of polymeric material so we expect this, combined with the work done on stretching the polymers, to give rise to strongly non-linear viscoelastic effects. This offers some contrast to the hypothesis that bacterial flagella on their own deplete their local environment of biopolymeric material and hence experience only the background Newtonian solvent \cite{Martinez2014}. 

\section{Conclusions}

Microbes live in complex fluidic environments, often of their own making. 
Microbial extracellular polymeric substances are continually secreted for a wide variety of purposes \cite{decho1990,wolfaardt1999}, including anchoring to surfaces by long mucous stalks \cite{Petroff2015,Petroff2014}, bioaccumulation of contaminants \cite{Wolfaardt1994}, and to serve as the polymeric matrix within biofilms \cite{Flemming2007}, veils \cite{Wirsen1978} and other collective structures \cite{schaudinn2007}. Motile swimmers must move through these complex media, while sessile microorganisms drive the transport of large high-conformational-entropy biopolymers. Previous work has focused on feeding currents \cite{higdon1979} entraining nutrients modelled as tracer-particles (which can be well described by hydrodynamic multipole expansion methods) \cite{mathijssen2015,mathijssen2016} or on the continuum limit of a viscoelastic fluid medium through which microbes must swim \cite{Lauga2007,Liu2011,Liu2014,Lauga2014,Patteson2016} or pump \cite{sleigh1988,VelezCordero2013} fluid. 
Both approaches average over the many internal degrees of freedom of long flexible biopolymers. 

We have shown how a rotating helical filament can accurately model the near-field fluid flow of a flagellated cell tethered to a wall, and studied the effects of this flow field on nearby polymers of various size. We have shown that long polymers are strongly attracted to the flagellum, and undergo a non-equilibrium stretching process as they are pulled towards it, and pumped along it. This implies that a dilute suspension of polymers tend to become locally concentrated in and around the flagellum rather than depleted.

Our results show that it is possible in simulations to measure the work applied to the polymer by the helix and that this is on average positive. Moreover, all of the work supplied to the polymer is dissipated, meaning that there is no elastic reclamation by the helix of the polymer's free energy when it collapses upon exit of the helix such as that observed in noiseless systems \cite{Ledesma-Aguilar2013,Wrobel2016}. Hence, our results provide some fundamental phenomenological insights into activity in microscopic, viscoelastic systems.

\textit{Acknowledgements} --- This work was supported through funding from the ERC Advanced Grant 291234 MiCE and we acknowledge EMBO funding to TNS (ALTF181-2013). AZ acknowledges funding by Marie Sk\l{}odowska Curie Intra-European Fellowship (G.A. No 653284) within Horizon 2020. We thank Alexander Petroff and Albert Libchaber for introducing us to veil-forming microbes. 

\bibliography{mergedRefs}

\end{document}